\documentclass[10pt,a4paper,aps,prd,showkeys,reprint,superscriptaddress]{revtex4-2}

\usepackage{graphicx,bm,mathptmx,amsmath,ulem}
\usepackage{natbib}
\usepackage{array}
\usepackage{xcolor}

\usepackage[breaklinks,colorlinks]{hyperref}

\usepackage{float}
\usepackage{enumitem}

\definecolor{prblue}{rgb}{0,0,.6}
\hypersetup{linkcolor=prblue,citecolor=prblue,filecolor=cyan,urlcolor=prblue,bookmarksnumbered=true,bookmarksopen=true}
\usepackage[all]{hypcap}

\newcommand\redsout{\bgroup\markoverwith{\textcolor{red}{\rule[0.5ex]{2pt}{1pt}}}\ULon}

\def\fauxschelper#1 #2\relax{%
  \fauxschelphelp#1\relax\relax%
  \if\relax#2\relax\else\ \fauxschelper#2\relax\fi%
}
\def\Hscale{.85}\def\Vscale{.74}\def\Cscale{1.12}
\def\fauxschelphelp#1#2\relax{%
  \ifnum`#1>``\ifnum`#1<`\{\scalebox{\Hscale}[\Vscale]{\uppercase{#1}}\else%
    \scalebox{\Cscale}[1]{#1}\fi\else\scalebox{\Cscale}[1]{#1}\fi%
  \ifx\relax#2\relax\else\fauxschelphelp#2\relax\fi}

\usepackage{textgreek} 
\usepackage{upgreek} 

\usepackage{catchfile}
\newcommand{\sh}[2][]{%
  \CatchFileEdef{\temp}{"|kpsewhich --var-value #2"}{\endlinechar=-1}%
  \if\relax\detokenize{#1}\relax\temp\else\let#1\temp\fi}

\newcommand{\xxx}[1]{}
\newcommand{\gc}{NGC~}

\newcommand{\cunits}{cm$^{-2}$}

\newcommand{\kmps}{km s$^{-1}$}

\newcommand{\lunits}{erg~s$^{-1}$}

\newcommand{\mbh}{$M_{\bullet}$}

\newcommand{\msun}{$M_{\odot}$}

\newcommand{\cobs}{$\cos\theta_{\rm obs}$}

\newcommand{\ewfeka}{$\rm EW_{Fe K\alpha}$}
\newcommand{\ewfekamc}{$\rm EW_{Fe K\alpha, MC}$}
\newcommand{\ewfekaapprox}{$\rm EW_{Fe K\alpha, approx}$}

\newcommand{\nh}{$N_{\rm H}$}

\newcommand{\tobs}{$\theta_{\rm obs}$}

\newcommand{\x}{X-ray}

\newcommand{\afe}{$A_{\rm Fe}$}

\newcommand{\fek}{Fe~K}
\newcommand{\feka}{Fe~K$\alpha$}

\newcommand{\fekb}{Fe~K$\beta$}

\newcommand{\feone}{Fe{\sc \,i}}

\newcommand{\chandra}{\textit{Chandra}}

\newcommand{\myt}{{\textsc{mytorus}}}

\newcommand{\cthick}{Compton-thick}
\newcommand{\er}{Equation~\ref}
\newcommand{\fr}{Figure~\ref}

\newcommand{\scr}{Sec.~\ref}

\newcommand{\exi}{\begin{equation}}
\newcommand{\exo}{\end{equation}}

\newcommand{\aer}[3]{$#1^{#2}_{#3}$} 
\newcommand{\ten}[2]{$#1\times 10^{#2}$} 
\newcommand{\up}[1]{$^{#1}$}

\def\spose#1{\hbox to 0pt{#1\hss}} 
\def\approxlt{\mathrel{\spose{\lower 3pt\hbox{$\sim$}}
        \raise 2.0pt\hbox{$<$}}}
\def\approxgt{\mathrel{\spose{\lower 3pt\hbox{$\sim$}}
        \raise 2.0pt\hbox{$>$}}}
\newcommand{\simlt}{$\approxlt$}
\newcommand{\simgt}{$\approxgt$}

\definecolor{aqua}{rgb}{0.0, 1.0, 1.0}

\newcommand{\sss}{$\sim$}

\usepackage{tikz}

\newcommand\aj{{AJ}}
\newcommand\araa{{ARA\&A}}
\renewcommand\apj{{ApJ}}
\newcommand\apjl{{ApJL}}     
\newcommand\apjs{{ApJS}}
\newcommand\aap{{A\&A}}
\newcommand\aapr{{A\&A~Rv}}

\newcommand\isci{{iScience}}
\newcommand\mnras{{MNRAS}}
\newcommand\pasp{{PASP}}
\renewcommand\nat{{Nature}}
\newcommand\gca{{GeCoA}}

\begin{document}

\title{The Ubiquity and Magnitude of Large FeK$\alpha$ Equivalent Widths in AGN Extended Regions} 

\author{P.~Tzanavaris}
\affiliation{Center for Space Science and Technology, University of Maryland, Baltimore County, 1000 Hilltop Circle, Baltimore, MD 21250}
\affiliation{Center for Research and Exploration in Space Science and Technology, NASA/Goddard Space Flight Center, Greenbelt, MD 20771}
\affiliation{The American Physical Society, Hauppauge, New York 11788}
  
\author{T.~Yaqoob}
\affiliation{Center for Space Science and Technology, University of Maryland, Baltimore County, 1000 Hilltop Circle, Baltimore, MD 21250}
\affiliation{Center for Research and Exploration in Space Science and Technology, NASA/Goddard Space Flight Center, Greenbelt, MD 20771}

\author{S.~LaMassa}
\affiliation{Space Telescope Science Institute, 3700 San
  Martin Drive, Baltimore, MD 21218}

\date{\today}

\begin{abstract}
\noindent Narrow \feka\ fluorescent emission lines arising at \sss
kpc-scale separations from the nucleus have only been detected in a
few AGN.  The detections require that the extended line emission be
spatially resolved and sufficiently bright.  Compared to narrow
\feka\ lines arising closer to the nucleus, they have much lower
fluxes but show substantially larger equivalent widths, \ewfeka. We
show that,
in the optically-thin limit,
a purely analytical argument
naturally predicts large, \ewfeka~\sss~1~keV, values for such lines,
regardless of the details of equivalent hydrogen column density, \nh,
or reprocessor geometry.
Monte Carlo simulations
corroborate this result and
show that the simple analytic \ewfeka\ prescription holds up to higher \nh\ approaching the Compton-thick regime.
We compare to
\chandra\ observations from the literature and discuss that our
results are consistent with the large \ewfeka\ values reported for
local AGN, for which the line is detected in extended, up to
$\sim$kpc-scale, regions.
We argue that large \ewfeka\ from kpc-scale regions in AGN should
  be ubiquitous, because they do not depend on the absolute luminosity
  of the central X-ray source, and are measured only against the
  scattered continuum. We predict values to be of the order of \sss1
  keV or larger, even for covering factors $\ll$1, and for arbitrarily
  small column densities.
We propose that
the large-scale molecular material that is now routinely being detected
with the Atacama Large Millimeter/Submillimeter Array (ALMA) may act
as an extended \x\ scattering reprocessor giving rise to \sss
kpc-scale \feka\ emission.

\end{abstract}


\maketitle

\section{Introduction}\label{sec-intro}

%

In the \x\ spectra of galaxies that harbor a nuclear actively
accreting, supermassive ($10^6$~\simlt~\mbh/\msun~\simlt~$10^9$) black
hole (SMBH), collectively known as Active Galactic Nuclei (AGN), the
spatial origin of the fluorescent, narrow (Full Width at Half Maximum,
FWHM~$<10\,000$~\kmps) \feka\ emission line at a rest energy of
6.4~keV, remains elusive.  This line is ubiquitous in both Type 1 and
Type 2 Seyfert galaxies and AGN\footnote{We use these terms
interchangeably in this paper.} with 2--10~keV luminosities
$<$$10^{45}$~\lunits. The line mean FWHM is
\sss2000~\kmps\ established from \chandra\ High Energy Transmission
Grating (HETG) spectra
\citep[][]{yaqoob2004,nandra2006,shu2010,shu2011}, although
\citep{liu2016} has suggested that the HETG line widths might actually
be over-estimated.  Although other, ionized Fe emission lines in the
\x\ regime are also reported in AGN, all observational evidence
strongly suggests that emission peaking at \sss 6.4~keV is the most
common Fe fluorescence feature in AGN \x\ spectra. The material in
which this line arises must then be neutral and relatively cool
\citep[][and references therein]{ghisellini1994}. Because the line is
narrow, it must be associated with distant matter at tens of thousands
of gravitational radii from the strong-gravity regime associated with
the central black hole. In this paper we are not concerned with broad
\feka\ line emission, which may also be observed in AGN and is a
manifestation of gravitational redshifting and Doppler broadening in
the strong-gravity regime. All reference to ``\feka\ emission'' and
``the line'' will imply the narrow line.

The spatial origin of the narrow line is thus often associated with
the putative obscuring, geometrically thick, dusty, molecular
``torus'' at a few parsecs from the SMBH. Regardless of the specific
details of the torus geometry and structure, it remains an essential
component of the AGN unification paradigm \citep[][see
  \citep{netzer2015,hickox2018,padovani2017} for
  reviews]{antonucci1993,urry1995}.  The distance from the SMBH and
size can be estimated directly from the narrow-line FWHM if the BH
mass is known.  This allowed \citep{shu2010,shu2011} to establish that
there is variation from object to object, with distances ranging from
the Broad Line Region (BLR) to the Narrow Line Region (NLR) Estimates
are also based on near- and mid-IR reverberation time lags
\citep[e.g.][]{burtscher2013,nunez2015,almeyda2020,figaredo2020,lyu2021},
assuming the \x\ torus is essentially the same as the IR torus.
Further, although \x\ \feka\ reverberation is mostly associated with a
{\it broad} \feka\ line
\citep[e.g.][]{fabian2009,zoghbi2010,zoghbi2013,kara2016,cackett2021},
      {\it narrow-line} reverberation results suggest that in a
      prominent AGN such as \gc4151 narrow \feka\ emission may
      originate in the inner BLR\footnote{This object was formerly
      also a famous candidate for {\it relativistic}
      \feka\ reverberation, but this is no longer the case; see
      \citep{zoghbi2019}.} \citep{zoghbi2019}.  In some sense the
      torus represents a transition region between the optical BLR
      closer to the nucleus and the optical NLR in the outer
      circumnuclear galactic environment.

Thus \feka\ emission origin in the BLR is also possible
\citep[e.g.][who report clumpy structures]{wang2017}, but also in the
region further out from the torus.  It is this spatially ``extended''
\feka\ line emission that we are concerned with in this paper, as
opposed to the usual \sss pc-scale narrow \feka\ emission closer to
the nucleus.  Molecular, geometrically thick obscuring material in
this region beyond the torus is reported e.g.~by~\citep{hicks2009} at
\sss30~pc, while \citep{prieto2014} \citep[see
  also][]{goulding2009,goulding2012} consider whether kpc-scale dust
filaments might be sufficient to account for all obscuration.
Notably, extended, specifically \feka\ emission, has been reported in
a few nearby AGN, in which the size scale could be spatially resolved
via \chandra\ CCD-imaging observations.  These are usually systems
estimated to be ``Compton thick,'' i.e.~with equivalent neutral
hydrogen column densities \nh~\simgt~\ten{1.25}{24}~\cunits, where the
Thomson optical depth becomes $>$1.  The flux of this extended line is
usually much lower than that of the usual line associated with the
torus, and as low as just a few percent of the total \feka\ emission
associated with a given object.  In order of increasing distance from
the nucleus, such emission is reported to originate up to \sss tens of
pc for Circinus \citep{marinucci2013}, hundreds of pc for \gc4945
\citep{marinucci2012,marinucci2017}, \sss300 pc for Mrk~3
\citep{guainazzi2012}\footnote{\citep{guainazzi2016} report a column
density in the Compton-thin regime.}, \sss1 kpc for ESO~428$-$G014
\citep{fabbiano2017}, and \sss2.2 kpc for \gc1068
\citep{young2001,bauer2015}.
Further, in \gc5643 \citep{fabbiano2018} an elongated north-south
\feka\ emission feature is identified over \sss65 pc
\citep{fabbiano2018}. In the case of \gc4388, thought to be a
``Compton-thin'' AGN \citep{yaqoob2023}, \citep{yi2021} stack \chandra\ ACIS-S data
from two observations and obtain significant detections of extended
\feka\ emission out to \sss0.8 kpc or \sss10~arcsec, most prominent in
three regions, labeled ``cones.'' They use
the disk-reflection continuum model of \citep{magdziarz1995} with a
Gaussian emission line to measure \ewfeka\ values of
\aer{474}{+71}{-70} eV and \aer{1.415}{+0.33}{-0.33} keV, for the
nucleus and the extended region, respectively. In addition, these
authors compile a sample of six AGN from the literature with spatially resolved
\feka\ extended emission and measured equivalent widths, \ewfeka,
which provides an extended emission \ewfeka\ baseline for comparative
studies. The measured \ewfeka\ values all fall in the
\sss1-2~keV range.

These consistently large \ewfeka\ values provide the motivation for
this paper, in which we use an analytical approximation to show that
these observed large \ewfeka\ values are to be expected in extended
AGN regions, regardless of the column density of the extended region,
even when the material is Compton-thin. We use Monte Carlo (MC)
simulations to calculate the extended region \ewfeka\ values for a
wide range of column densities, and intrinsic continuum slopes, and
show that the analytical approximation is useful for column densities
up to several factors of $10^{23}$ \cunits, a regime in which
line-emitting matter is optically-thin to scattering or absorption at
6.4~keV.

The structure of the paper is as follows: \scr{sec-cthin} introduces
the analytical approximation (\ref{sec-anal}) and presents the results
of MC simulations (\ref{sec-mc}).  \scr{sec-comp} compares our results
with published results from \chandra\ observations.  We discuss our
findings in \scr{sec-disc} and conclude with \scr{sec-summ} which
includes an overall summary.

\section{The \feka\ emission line EW in the Optically-Thin Limit}\label{sec-cthin}
\subsection{Analytical Calculation}\label{sec-anal}
We detail below how the \feka\ emission line equivalent width,
\ewfeka, can be obtained {\it analytically} in the optically-thin
limit.  This discussion is based on the very definition of equivalent
width, which is given by the line flux normalized by a continuum at
the line peak energy. The choice of continuum is usually what the
observer measures, which may consist of contributions from more than
one physically distinct regions in the source, if it is not spatially
resolved. Alternatively, EW values may be calculated with respect to
different continuum components obtained from modeling the net
spectrum.

\subsubsection{\feka\ line flux}
Following \citep{yaqoob2001}, we assume a uniform, spherical
distribution for the reprocessing material, with an \x\ point source
located at the center, and an incident power-law continuum $N_p
\ E^{-\Gamma}$ photons cm$^{-2}$ s$^{-1}$.  The line flux is
proportional to the number of continuum photons above the \fek\ edge
threshold, $E_{\rm K} \equiv 7.11$~keV for \feone, that are removed,
or
\begin{eqnarray}\label{equ-feflux-exact}
I_{\rm FeK\alpha} = f_c \ \omega_{\rm K} \ f_{\rm K \alpha} \int^{\infty}_{E_{\rm K}} N_p \ E^{-\Gamma} \left[ 1-\exp( -\sigma_{\rm FeK} \ A_{\rm Fe} \ N_{\rm H} ) \right] dE \nonumber\\
{\rm photons \ cm^{-2} \ \ s^{-1} },
\end{eqnarray}
where $f_c \equiv \Delta\Omega / 4\pi$ is the covering factor,
$\omega_{\rm K}$ the fluorescence yield for neutral Fe, $f_{\rm K
  \alpha}$ the fraction of emission-line photons appearing in the
\feka, and not the \fekb, line, $\sigma_{\rm FeK}(E)$ is the K-shell
photoelectric absorption cross-section, and \afe\ is the Fe abundance
relative to hydrogen.

We set $\sigma_{\rm FeK}(E) \equiv \sigma_0 (E/E_{\rm K})^{-\alpha}$.
We use $\alpha=2.67$ and $\sigma_0 =
3.37\times10^{-20}$\cunits\ from fits to Verner tables \citep[see
  also][]{murphy2009}.

It is important to note that in~\er{equ-feflux-exact} line photons,
once created, do not further interact with the reprocessing matter
  either by absorption or scattering.  In other words, the reprocessor
  is optically-thin ($\tau \ll 1$) to scattering and absorption at 6.4
  keV.  To linearly expand the exponential, we also impose optically
  thin conditions for the material to Fe-K absorption just above the
Fe~K edge, and thus also to all higher energies, since absorption
opacity decreases with energy.  In short, the optically-thin
condition, both to scattering and absorption, leads to photons
interacting with the material {\it at most once} for all energies
higher than 6.4~keV.

By expanding
the exponential, we then obtain the approximate relation
\begin{eqnarray}\label{equ-feflux}
I_{\rm FeK\alpha} \simeq f_c \ \omega_{\rm K} \ f_{\rm K \alpha} N_{\rm H} \ A_{\rm Fe} \sigma_0 \ E_K^{\alpha} N_p   \int^{\infty}_{E_{\rm K}} \ E^{-(\Gamma+\alpha)}  dE \nonumber\\
{\rm photons \ cm^{-2} \ \ s^{-1} } .
\end{eqnarray}

\subsubsection{\feka\ line-normalizing continuum}
In general, there are two main components to the continuum emission: The direct, unscattered continuum, consisting of source photons that are neither scattered nor absorbed; and the scattered continuum.
However, studies that report extended \feka\ emission
exclude by design emission from the
AGN nucleus, and there is no other direct hard \x\ emission from the
extended region. Only the scattered continuum is then of relevance for our
purposes.
The normalizing continuum,
due to photons scattered into the
line-of-sight by material with a Thomson
depth $\tau_{\rm sc}$, is thus given by
\begin{eqnarray}\label{equ-cont}
I_{\rm sc} & = & f_c \ N_P \ E^{-\Gamma}_0 \ (1-e^{-\tau_{\rm sc}}) \nonumber\\
& \simeq & f_c \ N_p \ E^{-\Gamma}_0 \ \tau_{\rm sc} \nonumber\\
& \simeq & f_c \ N_p \ E^{-\Gamma}_0  \ N_e  N_{\rm H} \ \sigma_T \ \nonumber\\
& &{\rm photons \ cm^{-2} \ \ s^{-1} \ keV^{-1}} ,
\end{eqnarray}
where $E_0 = 6.4008$~keV, the weighted average energy of
the centroids of the
Fe~K$\alpha_1$ and K$\alpha_2$ emission lines. $N_e$ is the number of
electrons per hydrogen atom. The energy is low enough that the
scattering cross-section is essentially the Thomson one, $\sigma_T$,
and the medium is optically thin to scattering
($\tau_{\rm sc} \ll 1$).
As in the previous section, the optically-thin limit implies
that after the first scattering a
continuum photon never interacts with the medium again, i.e.~the
photon escape probability is essentially unity because the medium is
optically thin to scattering and absorption at 6.4~keV.

It is worth pointing out that, since at lower energies the
  absorption opacity increases substantially, at some critical energy
  below 6.4~keV for a given \nh, the medium will no longer be in the
  optically-thin limit, and one would see absorption imprints on the
  scattered continuum. However, this does not affect our calculations
  and results, which do not involve these lower
  energies. Observationally, the scattered continuum may indeed show
  absorption signatures at low energies, and these could potentially
  be utilized to constrain modeling, provided the features are not
  too weak or swamped by other spectral features in the soft
  \x\ band.

\subsubsection{\feka\ EW}
We finally obtain an expression for the EW of the \feka\ line by
dividing \er{equ-feflux} by \er{equ-cont}:

\begin{eqnarray}\label{equ-ew}
{\rm EW}_{\rm FeK\alpha} = \omega_{\rm K} \ f_{\rm K \alpha} \ N_e \ A_{\rm Fe} \sigma_0 \sigma_T \ E_K^{\alpha} \ E_0^{\Gamma} \    \int^{\infty}_{E_{\rm K}} \ E^{-(\Gamma+\alpha)}  dE \nonumber\\
{\rm keV } \nonumber \\
\simeq 0.970 \ {\rm keV}
\frac{\omega_{\rm K}}{0.347} \ \frac{f_{\rm K \alpha}}{0.881} \ \frac{A_{\rm Fe}}{4.68\times10^{-5}} \frac{\sigma_0}{3.368\times10^{-20}} \\
\frac{3.57}{\Gamma+\alpha-1} \ (0.8985)^{(\Gamma-1.9)} 
\end{eqnarray}

We have assumed standard values for normalizing the constants in this expression
\citep[see][]{yaqoob2001}.
In addition, if the hydrogen and helium abundances are $A_{\rm
    H}$ and $A_{\rm He}$, respectively, the number of electrons per
  hydrogen atom is given by $N_e = (A_{\rm H} + 2 A_{\rm He}) / A_{\rm
    H} \simeq 1.22$ for the \citep{anders1989} abundances. While we
  do not assume a particular iron abundance, the value
  $4.68\times10^{-5}$ in the equation is the \citep{anders1989} value
  for solar Fe abundance.  

  This result, following directly from the imposed optically-thin
  limit, has a remarkable implication: The EW is independent of the
  covering factor and column density, with the implication that the EW
  is also independent of the detailed geometry, even though a
  spherical geometry was initially assumed.

\subsection{Monte Carlo Simulations}\label{sec-mc}
We now investigate the same question, namely the magnitude of
\ewfeka\ for pure reflection as a function of \nh, by adopting a
numerical approach.  We show that results from Monte Carlo (MC)
simulations of AGN \x\ reprocessing are entirely consistent with the
above analytic approximation, but in addition extend the analytic
result closer
to the \cthick\ regime.

Specifically, we probe the parameter space defined by \ewfeka, \nh,
intrinsic power law continuum index, $\Gamma$, and the cosine of the
angle between the torus symmetry axis and the observer, $\cos\theta$.
To this end, we use the MC results of ray-tracing simulations that
were performed to construct the model tables now incorporated in the
\myt\ model for \x\ spectral fitting.  Since we are interested in pure
reflection, the direct continuum is irrelevant for this analysis.

\begin{figure}
\hspace{-2cm}
\includegraphics[trim=60 0 0 20,clip,scale=0.7]{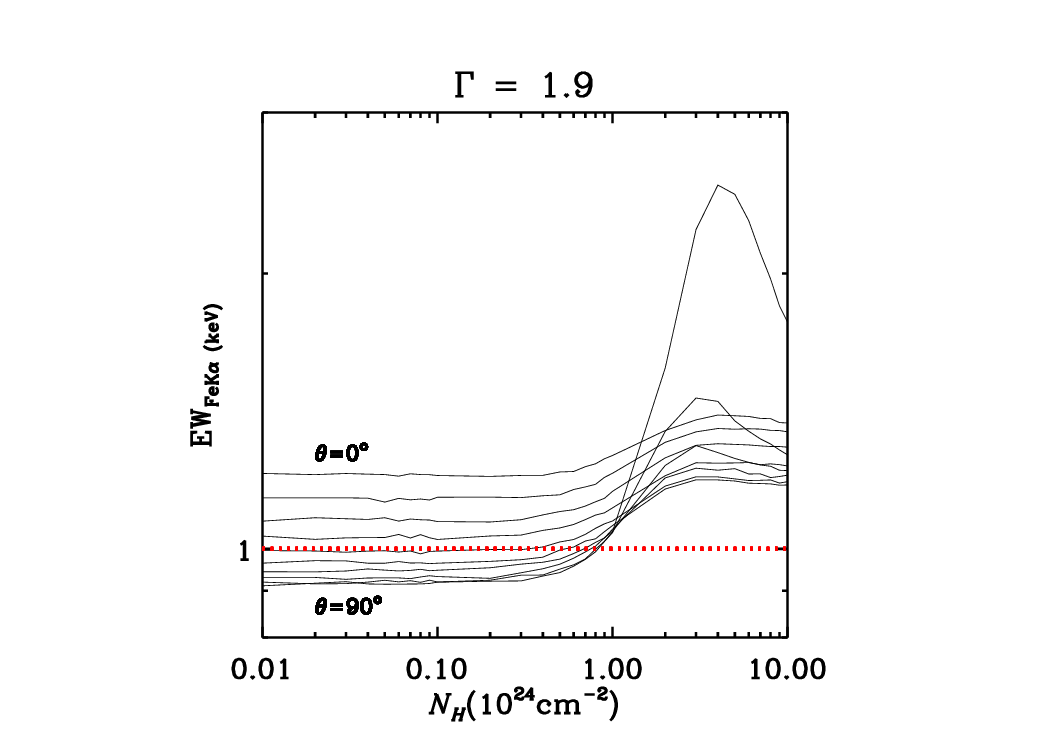}
\caption{\ewfeka\ as a function of \nh\ in
  \myt\ simulations. Different curves correspond to different
  cos$\theta$ values (or angle bins).  The red horizontal line shows
  the analytical result for the optically-thin limit (see text). Note
  that this is closest to the result for angle bin 5
  ($\cos\theta=0.5$), which corresponds to grazing incidence on the
  torus. Results are largely independent of \nh\ in the optically-thin and Compton-thin regime.}
\label{fig-ewnh-gamma1.9}
\end{figure}

We make use of the original MC simulations for the \myt\ model, which 
are described in detail in \citep{murphy2009} \citep[see also][]{yaqoob2012}. Briefly, these are 
simulations of Green's functions, covering
\nh\ values across the Thomson-thin to Compton-thick regime, for
incident photon energies up to 500 keV, and solar Fe abundance. The
reprocessed (``reflected'') continuum and its associated \feka/$\beta$ and
Ni K$\alpha$ emission are generated self-consistently with no ad hoc
components. Using the simulation output grid, we calculate
\ewfeka, i.e.~the equivalent width for the 0$^{\rm th}$ order \feka\ fluorescent line
(or more precisely the weighted centroid of \feka1 and \feka2) relative to the
Compton scattered continuum as a function of:

\begin{enumerate}[noitemsep]
\item the cosines of the centers
  of 10 angle bins, \cobs, corresponding to line-of-sight angles from
  \tobs~$=0^{\circ}$ (bin 1) to $=90^{\circ}$ (bin 10);
\item 13 values of the intrinsic incident power-law index from
  $\Gamma=1.4$ to 2.6;
  \item 28 values of equatorial equivalent hydrogen column density from \nh~$=0.01$ to $10 \ (\times10^{24}$\cunits).
\end{enumerate}

\begin{figure}
\hspace{-2cm}
\includegraphics[trim=60 0 0 20,clip,scale=0.7]{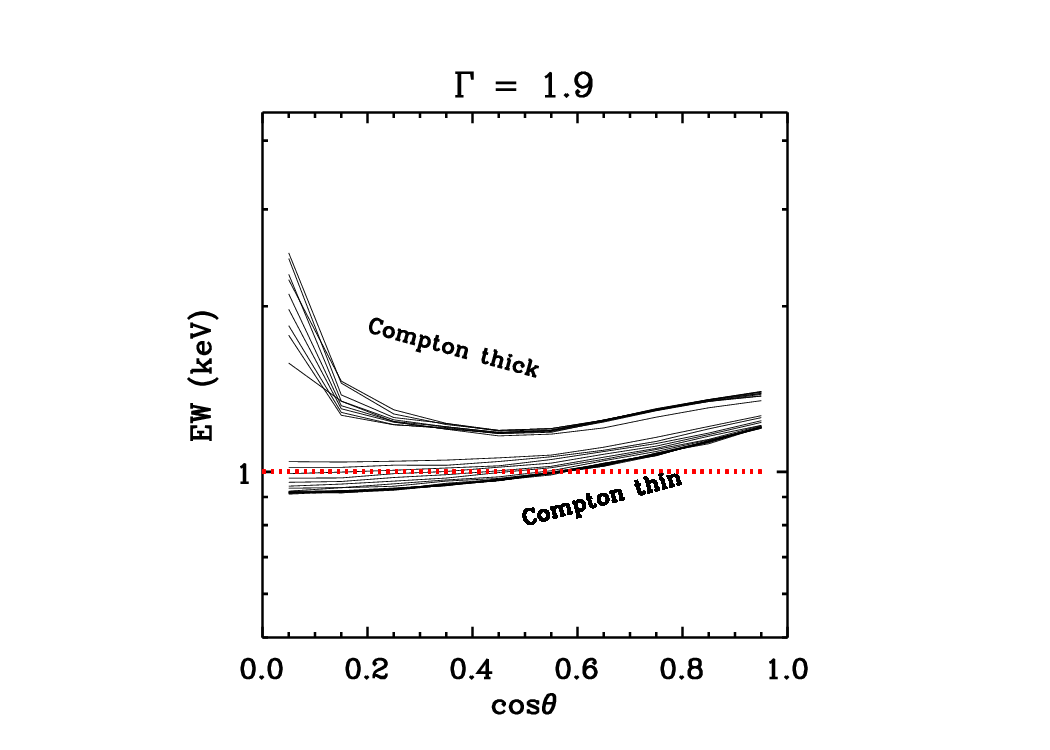}
\caption{\feka\ EW as a function of $\cos\theta$ (or angle bin) in
  \myt\ simulations. Different curves correspond to different
  \nh\ values in the Compton-thin (lower group of curves, $10^{22}$ to
  $10^{24}$~\cunits) and Compton-thick (upper group of curves,
  $2\times10^{24}$ to $10^{25}$~\cunits.) regime.  The red horizontal
  line shows the analytical result for the optically-thin limit (see
    text). Note that this is closest to the result for bin 5
    ($\cos\theta=0.5$), which corresponds to grazing incidence on the
    torus. Results are mostly independent of \nh\ in the optically-thin and Compton-thin regimes.}
\label{fig-ewcos}
\end{figure}

We show the simulation-based dependence of \ewfeka\ on \nh\ for
$\Gamma=1.9$ and all angle bins in \fr{fig-ewnh-gamma1.9}. In a given angle bin,
there appears to be no dependence on \nh\ up to
$\sim$\ten{4}{23}~\cunits. This can also be seen in \fr{fig-ewcos},
which plots \ewfeka\ against $\cos\theta$. Here, there are two distinct
groups of curves: The lower group corresponds to \nh\ values from
$10^{22}$ (lowest curve) to $10^{24}$~\cunits\ (topmost curve). The
upper group of curves corresponds to \nh\ values in the Compton-thick
regime, from \ten{2}{24} to $10^{25}$~\cunits. In both Figures, the
analytical result of the previous Section is overplotted as a dotted
red line, and clearly agrees best with the MC result for bin 5
($\cos\theta=0.5$, $\theta=60^{\circ}$).

\begin{figure}
\hspace{-2cm}
\includegraphics[trim=60 0 0 20,clip,scale=0.7]{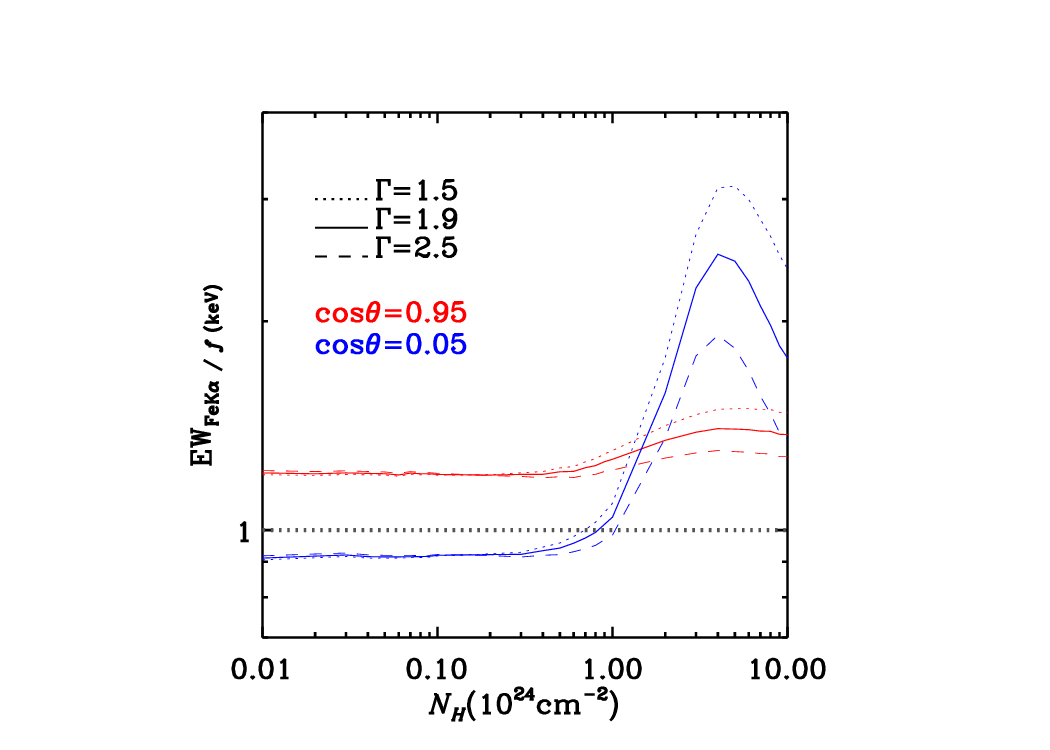}
\caption{\feka\ EW normalized by a factor $f(\Gamma)$ (see text) as a
  function of \nh\ for different values of $\cos\theta$ and
  $\Gamma$. The two extreme $\cos\theta$ values of 0.95 (face-on, red
  upper curves) and 0.05 (edge-on, blue lower curves) form an envelope
  enclosing intermediate results (not shown for clarity). Three
  different $\Gamma$ values are shown by a dotted, dashed, and solid
  curve in each $\cos\theta$ case. The analytical optically-thin
    limit (\er{equ-ew}) is shown by the grey dotted horizontal
    line. Results are clearly independent of \nh\ up to $\sim
    3-5\times 10^{23}$\cunits, with the exact threshold of dependence
    slightly depending on $\theta$ and $\Gamma$.}
\label{fig-ewFnh-gammas}
\end{figure}

It can be seen in \fr{fig-ewnh-gamma1.9} that there are two extreme cases for
$\theta=0^{\circ}$ and $\theta=90^{\circ}$, effectively defining an
``envelope'' in $\theta$ (and $\cos\theta$).  In \fr{fig-ewFnh-gammas}
we show the dependence of \ewfeka\ on $\cos\theta$ for three
characteristic $\Gamma$ values covering a plausible range between
$1.5$ and $2.5$.  Here, \ewfeka\ is normalized by the factor $f \equiv
\frac{3.57}{\Gamma+\alpha-1} \ (0.8985)^{(\Gamma-1.9)}$ (see
\er{equ-ew}), thus removing the explicit dependence on $\Gamma$.
Note that $f\simeq1$ for $\Gamma=1.9$ (and given that $\alpha=2.67$).

\begin{figure}
\hspace{1cm}
\includegraphics[trim=40 0 0 10,clip,scale=0.73]{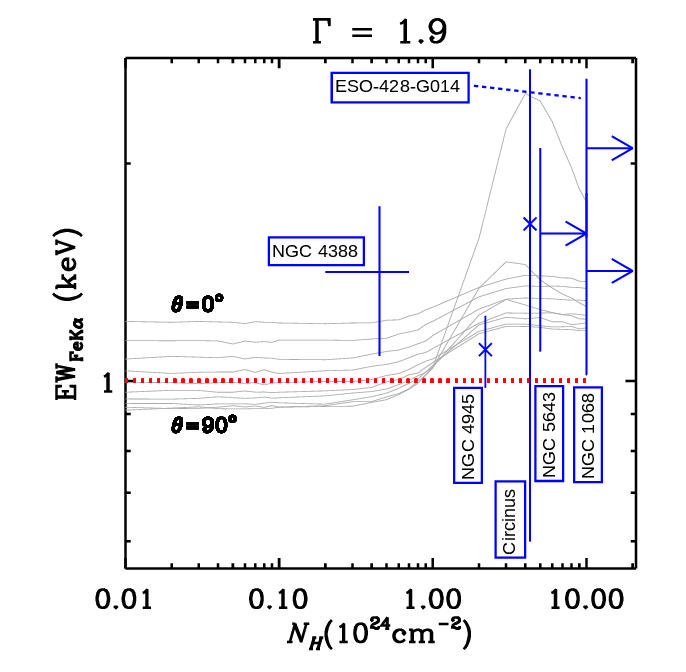}
\caption{Same as \fr{fig-ewnh-gamma1.9} with data for the six AGN with detected \feka\ extended emission from the compilation of \cite[Table 4]{yi2021} as indicated. The right-pointing arrows represent lower limits. Otherwise, blue line segments represent the range of estimated values, taking into account both ranges of measurements and reported uncertainties, where available.}
\label{fig-ewnh-data-gamma1.9-names}
\end{figure}


\section{Comparison with \textbf{\textit{Chandra}} Observations}\label{sec-comp}

\fr{fig-ewnh-data-gamma1.9-names} is a modified version of
\fr{fig-ewnh-gamma1.9}, with the \chandra-based compilation of results
for detected extended \feka\ emission presented in \citep[][Table
  4]{yi2021} overplotted. As explained by these authors, the reported
results for Circinus cover a range of earlier results in \ewfeka. For
clarity, we show here the full range in \ewfeka\ for this object, taking into account
uncertainties. The average central value is shown with a cross. For
the remaining systems, we show single \ewfeka\ central values with uncertainties
from the references reported in \citep{yi2021}. Observationally
estimated \nh\ values compiled by \citep{yi2021} do not have uncertainties. For \gc4388 the horizontal ``error bar'' represents 
the range of reported \nh\ central values. Three other AGN have
lower limits in \nh\ as indicated by the arrows.

Focusing on \ewfeka\ values shown in \fr{fig-ewnh-data-gamma1.9-names},
we note that all reported values are within a factor of \sss2
of the analytic approximation of \sss1~keV. In the case of
Compton-thin \gc4388 in particular, which is most consistent
with the assumptions of the analytic approximation, \ewfeka\ is also
close to the analytic value within the reported \ewfeka\ errors.
Finally, all \ewfeka\ values are consistent with the MC values
within the reported uncertainties and ranges.

\section{Results and Discussion}\label{sec-disc}

A key result from this work is that the optically-thin analytic
approximation for \ewfeka\ is surprisingly close to the MC results
for column densities that go well beyond the optically-thin regime, up
to \nh~\simlt~\ten{4}{23}~\cunits. In this regime, \ewfeka\ is only weakly
dependent on $\theta$.  We discuss these results further below.

\begin{figure}
\hspace{-2cm}
\includegraphics[trim=60 0 0 20,clip,scale=0.7]{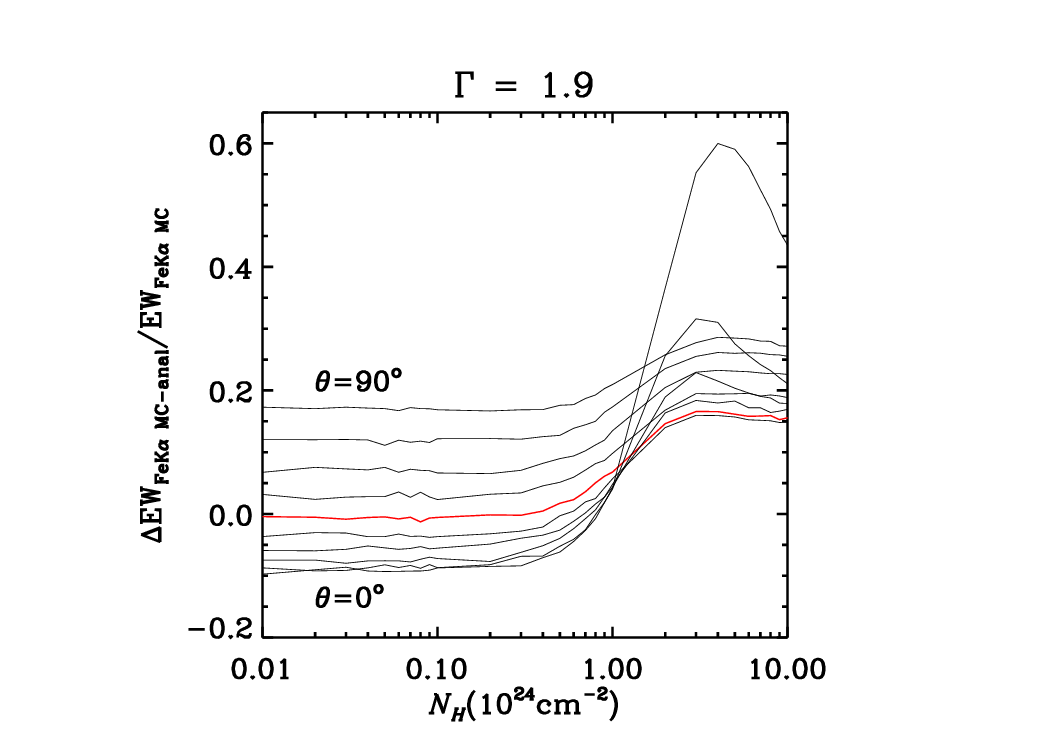}
\caption{As \fr{fig-ewnh-gamma1.9} but for the {\it fractional}
  difference between \ewfeka\ from MC simulations compared to the
  optically-thin limit (analytic result in \er{equ-ew}) as a function of
  \nh. The red curve corresponds to angle bin 5 ($\cos\theta=0.5$) in
  the simulations.}
\label{fig-ewnhfrac-gamma1.9}
\end{figure}

\begin{figure}
\hspace{-2cm}
\includegraphics[trim=60 0 0 20,clip,scale=0.7]{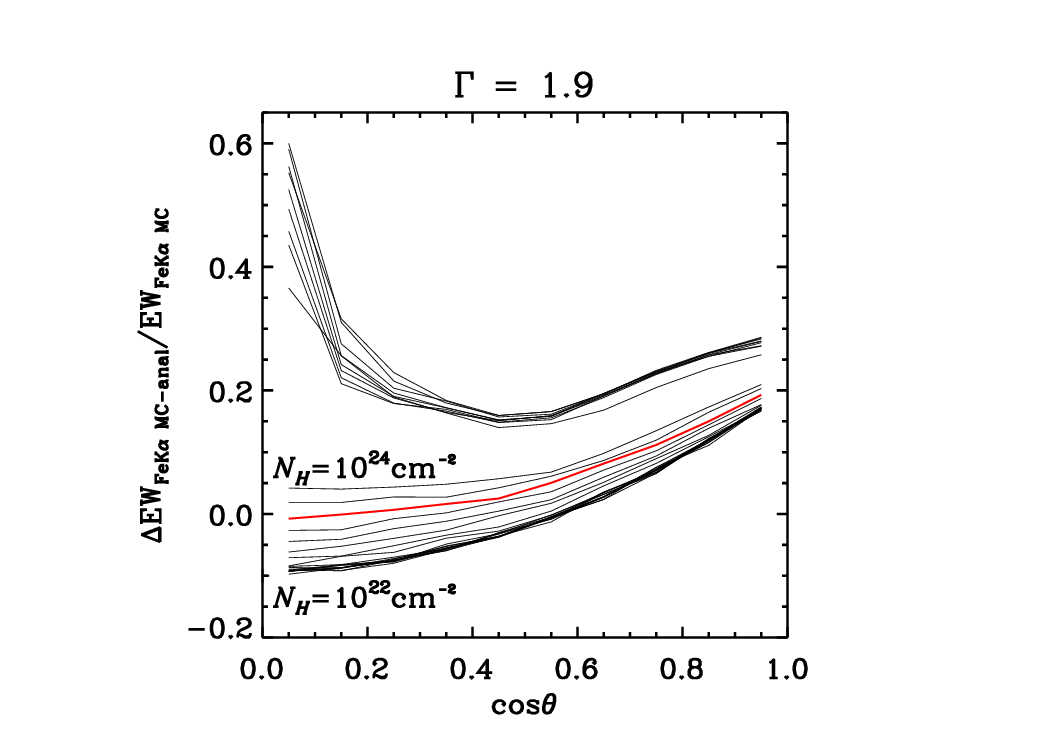}
\caption{As \fr{fig-ewcos} but for the {\it fractional} difference
  between \ewfeka\ from MC simulations compared to the analytic result
  (\er{equ-ew}, optically-thin limit) as a function of
  $\cos\theta$. The red curve corresponds to the MC simulation bin for
  \nh~$=8\times 10^{23}$~\cunits.  }
\label{fig-ewcosfrac-gamma1.9}
\end{figure}

Figures~\ref{fig-ewnh-gamma1.9} and \ref{fig-ewcos} show that the
\ewfeka\ estimated analytically is within \simlt20\%\ for {\it all} MC
estimates, regardless of $\theta$ bin, and up to
\nh~\sss~\ten{4}{23}~\cunits.  This is highlighted in the fractional
difference versions of the figures,
i.e.~Figures~\ref{fig-ewnhfrac-gamma1.9} and
\ref{fig-ewcosfrac-gamma1.9}.  For $\theta=60^{\circ}$ the agreement
is within \sss1\%, and even at \sss10\up{24}~\cunits\ it is within
\sss5\%\ (red curve in \fr{fig-ewnhfrac-gamma1.9}).
The
optically-thin
analytic approximation for \ewfeka\ was derived assuming a
spherical geometry,
which necessarily has material
intercepting the line of sight,
and it aligns most closely with the MC angular bin that has
the smallest non-zero column density. This is the grazing-incidence
angle bin, which has its bin boundary at $\cos\theta=0.5$, corresponding
to $\theta=60^{\circ}$. The analytic
\ewfeka\ approximation does not agree as well with MC results for
other angle bins, in particular those not intercepting any material,
even for small column densities, because it assumes non-zero columns
of material in the line of sight. Even so, an agreement within
\sss20\%\ for all bins is a significant result.

Further, \fr{fig-ewFnh-gammas} shows that curves of \ewfeka\ as a
  function of \nh\ diverge for different $\Gamma$ values above
  \nh\sss\ten{4}{23}~\cunits, implying that the $\Gamma$ dependence of
  \ewfeka\ in the analytic approximation fails to capture the physics
  above this column density. Thus, the analytic approximation breaks
  down above \nh\sss\ten{4}{23}~\cunits, and this column density
  represents the approximate upper limit of applicability of the
  analytic approximation. This column density corresponds to a Thomson
  depth of 0.32.

One might ask whether the toroidal geometry which is assumed in
  the \myt\ simulations might reduce the generality of these results.
  In the optically-thin regime, the agreement with the analytical
  approximation, which does not assume such a geometry, is one
  indication that this should not be an issue.  In the Compton-thick
  regime, we note the independent MC results of \citep{leahy1993}, who assume
  a uniform spherical geometry, obtaining
  \ewfeka~\simgt~1~keV that increases with optical depth / column density. Even though these authors do not exclude the direct continuum in their
  \ewfeka\ calculations as we do, excluding the direct continuum would only
  further {\it enhance} the large \ewfeka\ effect, as we oberve. Finally, even if the toroidal geometry were to have some effect, we still see that regardless of angle $\theta$, \ewfeka\ remains $>$1~keV for all \nh\ probed. This strongly suggests that the dominant effect is the exclusion of the direct continuum.

In the case of a clumpy geometry, each clump will produce a large
  \ewfeka\ as, once more, this would be measured relative to the
  scattered continuum only. The ensemble of clumps would then give
  rise to an overall large \ewfeka. In the limit of a high-filling
  factor, the results would be as for a sphere, discussed above
  \citep[see][]{leahy1993}, and in the optically thin limit the
  analytic approximation would once again hold.

Overall, given the simple assumptions underlying the analytic
approximation, and the fact that one might expect it to fail
significantly above 10\up{22}~\cunits, this is a significant result,
that provides a simple explanation for large equivalent widths of
\feka\ emission lines in extended AGN regions.

\citep{yi2021} note that \ewfeka\ values are \sss3
times larger in the extended region compared to those for regions
closer to the nucleus. They attribute this to differences in geometry or Fe
abundance between the circumnuclear and the extended region.
However, it should be pointed out that a major issue
  with such an explanation is that increasing the Fe abundance
  does not linearly increase \ewfeka\ because more Fe also means more
  absorption (including of line photons), and not just K- but also
  L-shell absorption. To increase \ewfeka\ by a factor of \sss3, you
  need {\it at least an order of magnitude} increase in Fe abundance
  \citep[see][Fig.~17]{george1991}, by which point
  the continuum will have become completely skewed and wrong.
Further,
as also explained by \citep[][see also references therein]{yi2021}, an
increase in iron abundance would mainly be introduced via delayed SN
Ia enrichment over timescales of \sss1 Gyr, but this would be unlikely to remain
preferentially in the extended circumnuclear region over such prolonged periods
of time.
Differences in geometry would imply that somehow more \x\ reprocessing
material would be located at larger scales compared to those usually
attributed to torus-like structures, which are thought to be up to few
pc based on virial assumptions for the \feka\ line width,
IR reverberation mapping, and ALMA sub-mm imaging \citep[e.g.][]{shu2010,gandhi2015,garcia-burillo2016,garcia-burillo2019,honig2019}.
\citep{yi2021} further consider the possibility of an intrinsically
depressed AGN continuum which would naturally favor larger EW
measurements. This however should also affect the EW measured for
\feka\ emission originating closer to the central AGN.

Instead, qualitatively \ewfeka\ in the extended region would
naturally be expected to be larger than that for the more common
narrow \feka\ line arising closer to the nucleus, simply because {\it
  the extended-emission line equivalent width is measured only with
  respect to the scattered continuum}, which is not
the case for the more common line.
This is a key point, and to our knowledge, no previous work on
  \ewfeka\ has appropriately taken this into account. In this
paper we are highlighting and quantifying this effect both
analytically and computationally.  Both approaches corroborate the
qualitative expectation, and are also in agreement with each other.

Both of the analytical expressions for the line flux
(Equations~\ref{equ-feflux-exact} and \ref{equ-feflux}), as well as
the one for the normalizing scattered continuum (\er{equ-cont}),
assume that the reprocessing matter is optically thin to both absorption and
scattering. Line photons effectively do not interact with this matter after being
created; similarly, continuum photons never interact with the medium
after the first scattering (either by absorption or further
scattering). Put differentely, in both cases, we are setting the photon escape
probability function to unity, which is justified in this regime. In
general, the \ewfeka\ expression (\er{equ-ew}) should include such an
escape function both in the numerator and the denominator. These
functions will be different in general, but when the medium is
optically thin, one can reasonably assume that the spatial distribution
of line creation sites and that of scattering sites are
the same because both are distributed uniformly in the medium.
However, as the optical depth increases, these distributions will not
remain the same, and the two escape functions become different.
What we have effectively done is to use the MC results as a computational
experiment to probe the evolution of these escape functions. As we
have shown, the reasonable, qualitative assumption that the
spatial distributions of the line creation and continuum scattering sites
are similar holds up to column densities \nh~\sss~$4\times10^{23}$~\cunits.
Therefore, they cancel out in the analytical \ewfeka\ expression, thus
making \ewfeka\ independent of \nh.

As to the actual nature of the extended emission material,
we note that cold, molecular material due to outflows in AGN at
larger, tens to hundreds of pc, scales has also been detected in the
sub-mm
\citep{curran2008,zschaechner2016,gallimore2016,alatalo2011,bolatto2021}. Recently, detections of molecular tori with extended
diameter sizes up to \sss50~pc are reported
\citep{combes2019,garcia-burillo2021} extending the obscuring torus
itself beyond the pc-scale paradigm. The proposed combined emerging
molecular and IR picture includes both outflows and feeding inflows
from resonant molecular reservoirs at \sss100~pc \citep{ramos-almeida2017,alonso-herrero2018,izumi2018,alonso-herrero2019,alonso-herrero2021,combes2019,honig2019} or more.  Overall, such
molecular material would also be a candidate for an extended {\it X-ray}
reprocessor.

It  is clear that even AGN with moderate levels of
obscuration, such as \gc4388 and Mrk~3, do show extended
\feka\ emission over spatial scales of at least hundreds of parsecs,
with associated keV-scale EWs.  
Our MC results have shown that a large
\ewfeka\ would be naturally expected
{\it for all} \nh.
Since \gc4388 is Compton-thin, its large \ewfeka\ should also be
better predicted by the analytic approximation
(which, however,
is based on the stricter optically-thin
condition)
and the analytical result of
\sss1 keV for \ewfeka\ is in good agreement with the lower limit of
the value reported by the \chandra\ analysis, i.e.~1.085 keV.  The
analytical result is also entirely consistent with \ewfeka\ values of
0.7$-$1 keV reported as due to molecular clouds around Sgr A\up{\ast}
scattering X-ray emission from nuclear flares
\citep{ponti2010},
although
these clouds are thought to be located tens rather than hundreds of
parsecs away from the nucleus.

Thus, overall, both the MC results and the analytical
  approximation suggest that large \ewfeka\ signatures should
  be ubiquitous at kpc-scale distances from AGN, and the few available
  observational results are in support of this picture.



\section{Summary and Conclusions}\label{sec-summ}
We have shown that the narrow \feka\ emission equivalent width
observed at \sss kpc scales in AGN can be predicted both analytically
and numerically, and have compared these predictions to
observed results from the literature. Our main conclusions are:
\vspace{-.3cm}
\begin{enumerate}
  \itemsep-4pt
  
\item
Calculations of \ewfeka\ in the optically-thin limit, coupled with the absence of the direct X-ray AGN continuum from the extended region,
   lead to an analytic approximate
  estimate \ewfekaapprox~\sss~1~keV that is independent of \nh\ or
  geometric details such as covering factor.
  
\item Using state-of-the-art MC ray-tracing simulations
with \myt, we
  show that \ewfekamc:

\vspace{-.3cm}
  \begin{enumerate}
    \itemsep-3pt
  \item is independent of \nh\ up to \sss4~$\times 10^{23}$~\cunits;
  \item is mildly dependent on the angle to the line-of-sight, $\theta$, and the power law index, $\Gamma$;
  \item is within \sss20\%\ (\sss1\%) of \ewfekaapprox\ for all
    $\theta$ (for $\theta=60^{\circ}$) up to \sss4~$\times
    10^{23}$~\cunits;
  \item is consistently $>1$~keV as the \nh\ increases into the Compton-thick \nh\ regime, suggesting that large \ewfeka\ values are to be expected {\it for all} \nh.
  
  \end{enumerate}

\item We argue that these results should remain unaffected by
  toroidal, spherical, or clumpy geometries. However, the results do
  not carry over to the absolute flux of the \feka\ line (as opposed
  to the EW): \citep{yaqoob2010} showed that for line flux, the
  optically-thin approximation breaks down at a column density of only
  $\sim 4 \times 10^{22} \ \rm cm^{-2}$.
  
\item Both \ewfekamc\ and \ewfekaapprox\ are within a factor of \sss2 from
  observational estimates for \ewfeka\ at \sss kpc scales in local AGN.

\end{enumerate}

The \ewfekamc\ and \ewfekaapprox\ good agreement up to
\nh~\sss~\ten{4}{23}~\cunits\ directly demonstrates
and quantifies a reasonable
expectation in the
optically-thin
regime.
Beyond this, the MC results show that \ewfeka\ will
  remain larger than 1~keV into the Compton-thick regime,
  a prediction that should be tested further with more observational data.
This agreement, as well as the order-of-magnitude agreement with
observational results, suggest that the relative prevalence of narrow
\feka\ AGN emission at kpc scales beyond the ``canonical'' torus
follows a roughly predictable pattern across AGN and at least two
orders of magnitude in \nh. Larger AGN samples with such detections,
as well as multiwavelength detections in the IR and sub-mm, would
provide further insight into the nature and frequency of large-scale
\x\ AGN reflection.
It also remains to be explored whether similar \ewfeka\
  behavior can be established at the smaller spatial scales
  of Galactic \x\ binaries.

\begin{acknowledgments}

\noindent
We thank the anonymous referee for their constructive comments that
helped improve this paper.  P.T. acknowledges support from NASA grants
80NSSC18K0408 (solicitation NNH17ZDA001N-ADAP) and 80NSSC22K0411
(solicitation NNH21ZDA001N-ADAP).  This work is supported by NASA
under the CRESST Cooperative Agreement, award number 80GSFC21M0002.

\end{acknowledgments}



\bibliographystyle{apsrev4-2}
%

\end{document}